# How does climate influences regional stability

**Tianyu Shi[a][1][1], Jiayan Guo[b][1], Xuxin Cheng[c], Yu Hao***

[a]School of Mechanical engineering, Beijing Institute of Technology, Beijing, China
[b]School of Computing, Beijing Institute of Technology, Beijing, China
[c]School of Automation, Beijing Institute of Technology, Beijing, China

**Abstract**

Regional stability is the key development concern for most part of the world, though such stability largely depends on region's economic condition. It is also affected by other natural factors such as climate change. In this paper, we analyze the influence of climate change on regional stability. We first used Principle Components Analysis (PCA) to extract key factors, and then established a BP neural network to evaluate these key components. The result indicates that annual precipitation has the highest correlation with regional stability among all climatic factors in the long run.

___________________________________________________________________________________________

## 1. Introduction

Global warming is now leading to multiple environmental effects including rising ocean temperatures, increasing air humidity over the ocean and more intense ocean storms. Recently, occurrence patterns of these effects are becoming more elusive to find out[1]. In general, such changes lead to more intense marine hurricanes and flood, which can count as disaster for human society. Such economic loss finally catches people's awareness. However, there is no easy solution to this global climate issue and the fixing cost is huge according to The Intergovernmental Panel on Climate Change[2].
Besides direct loss, such environmental changes may have further influence on human society in a negative way. Government trembles and chaos surfaces[3]. However, such impression is more of intuition and the real impact of climate change on regional stability is still vague. In early days, scientists have been researching on human activities' impact on regional stability. However, It is until very recently, that people started to look into other factors such as climate change that may have impact on regional stability[4]. Experts believe that in the future climate change will be more intense. For instance, extremely strong hurricanes will occur more frequently and make more landfalls[5]. And there come discussions in the science community about how these climate changes will affect regional stability in the future[6]. Although there has been some researches on the relationship between climate change and regional stability from qualitative perspective[3], the exact relationship still remains unclear.

Therefore it is of great value to study the relationship between climate change and regional stability from a quantitative perspective, which is the main purpose of this article.

## 2. Materials and methods

### 2.1 Data processing

#### 2.1.1 Introduction to PCA Analysis

In this section, we introduce PCA as a projection method, which gives information about the latent structure of the data set [7]. It uses an orthogonal transformation to transform a set of observations of potentially correlated variables into a set of values of linearly uncorrelated variables called principal components[8]. We first choose some variables which have potential correlation as influence factors of regional stability. For example, we choose flood occurrence frequency and drought occurrence frequency as variables, which are both influenced by the precipitation of a region and thus have correlation.

___________________________________________________________________________________________

[1] These authors contributed equally and should be considered co-first authors. *: Corresponding author.

Then we use Projection pursuit to assign the weight of indexes of principal components from the principal analysis. Projection pursuit seeks one projection at a time in order to ensure the extracted signal as non-Gaussian as possible. The idea of projection pursuit is to locate a projection or projections from high-dimensional space to low-dimensional space that contains the most details about the structure of the data set. Once an interesting set of projections has been found, existing structures (clusters, surfaces, etc.)[9] can be extracted and analyzed separately. Hence, weight of indexes of principal components from the principal analysis can be assigned via projection pursuit.

### 2.1.2 The Selection of Indexes

There are numerous factors influencing the stability of a region. The impact of climate change is one of them. To further view the best factors, we propose seven indexes indicating a country's stability first. These indexes include climate variables which present the region's climate features and government variables which indicate the government's ability to control the region. These indexes are shown in **Table 1**.

|  |  |
|---|---|
|  | Long Term Average Precipitation(LAP)[10] |
| Climate Indexes | Annual Average Temperature(AAT) [11] |
|  | Flood Occurrence(FO) [12] |
|  | Drought Occurrence(DO) [13] |
|  | Annual Military Spending(AMS) [14] |
| Government Indexes | Law Level(LL) [15] |
|  | Public Support Rate(PSR) [16] |

**Table 1.** Initial Indexes of Regional Stability. **LAP** and **AAT** have a lot to do with agriculture situation of the region. Proper **LAP** and **AAT** mean higher crop yields. **FO** and **DO** indicate the occurrence chance of extreme climate, which results in reduced crop yields and poor living conditions of local residents[17]. As for the Government Indexes, the **AMS** reflects the ability of the government to control the region and exterminate latent destabilizing factors. **LL** is the Law Level of the region and high law level means more stable living condition of the country. The last index is PSR, which reflects the credibility of the government. In conclusion, a good government will have relatively higher values of the three indexes above.

### 2.1.3 Data Collection

To make further research, Data of these indexes are collected and are presented in **Figure 1**. They are mainly collected from the world bank database which covers 179 countries and ranges from 1990 to 2016. We first preprocess the data to eliminate the abnormal data points. To make the data more intuitive, we provide the relative indexes rate between Sudan, Somalia and Yemen in 2016 shown in **Figure 1.**

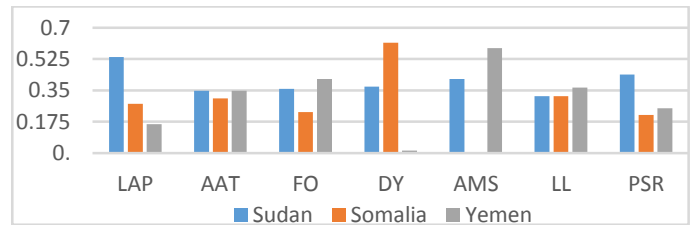

Through this figure, we can recognize that the situations in each country is complex. And it is not plain to see the most relative index directly. To make a break, we then implement the Principal Component Analysis to choose the top-k prime indexes to make further use.

### 2.1.4 Implement Process of PCA

As shown in **Figure 1**, the data have potential correlations among these columns. To construct a reasonable evaluation system of Region Stability, we should first transfer these correlated indexes into principal components. The core process of PCA is elaborated as follows.

**Step 1:** Construct a data matrix X:
$$X = \begin{bmatrix} X_{11} & \cdots & X_{1p} \\ \vdots & \ddots & \vdots \\ X_{n1} & \cdots & X_{np} \end{bmatrix}$$
where each row represents data of indexes of one country and each column gives a particular index of every country.

**Step 2:** Due to the different dimensions of these initial indexes, the data shall be standardized with the following formula:

$$x_{ij}^* = \frac{x_{ij} - \bar{x}_j}{\sqrt{Var(x_j)}}, for\ all\ i\ and\ j$$

where, $\bar{x}_j = \frac{1}{n}\sum_{i=1}^{n} x_{ij}$; $Var(x_j) = \frac{1}{n-1}\sum_{i=1}^{n}(x_{ij} - \bar{x}_j)^2$ for all j

**Step 3:** Calculate the sample correlation coefficient matrix:

$$R = \begin{bmatrix} r_{11} & \cdots & r_{1p} \\ \vdots & \ddots & \vdots \\ r_{p1} & \cdots & r_{pp} \end{bmatrix}$$

where, $r_{ij} = cov(x_i, x_j) = \frac{\sum_{i=1}^{n}(x_i - \bar{x}_i)(x_j - \bar{x}_j)}{n-1}, n > 1$

**Step 4:** Calculate the eigenvalues and the corresponding eigenvectors of the correlation coefficient matrix:
The eigenvalues: $\lambda_1, \lambda_2, \ldots, \lambda_p$
The eigenvectors: $v_i = (v_{i1}, v_{i2}, \ldots, v_{ip})$, for all i.

**Step 5:** The final principal components can be acquired from the Principal Component Analysis. Based on the calculated contribution rate of the principal components, the first k principal components are chosen. Cr is the contribution rate, which refers to the proportion of the variance of a principal component to the total variance:

$$Cr = \frac{\lambda_i}{\sum_{i=1}^{p} \lambda_i}$$

The higher contribution rate a principal component has, the more information the original variable contains.

**Step 6:** Using MATLAB to compute Cr, the results are shown in the **Table 2**.

| Initial Indexes | Eigenvalues | Cr | Accumulated Cr |
|---|---|---|---|
| LAP | 3.7366 | 41.14% | 41.14% |
| AAT | 1.8172 | 20.01% | 61.15% |
| AMS | 1.2306 | 13.55% | 74.70% |
| FO | 1.1533 | 12.70% | 87.40% |
| PSR | 0.7351 | 8.090% | 95.49% |
| LL | 0.3561 | 3.920% | 99.41% |
| DF | 0.0533 | 0.590% | 100.00% |

**Table 2.** The Calculated Contribution Rate of Indexes. Assuming that the calculated contribution rate is 95%, the decision of k principal components can be made. As is illustrated in **Table 2**, the first four principal components can be chosen, the calculated contribution rate of which is 95.49%. The ultimate principal components include Long Term Average Precipitation, Annual Average Temperature, Annual Military Spending, Flood Occurrence and Public Support Rate.

*2.2 Model building*

*2.2.1 The BP Neural Network of indexes*

We use the Back Propagation Neural Network[18] to finally describe Region Stability. BP Network is a three layer full-connecting artificial neural network containing input layer, hidden layer and output layer, which is described in the **Figure 2**.

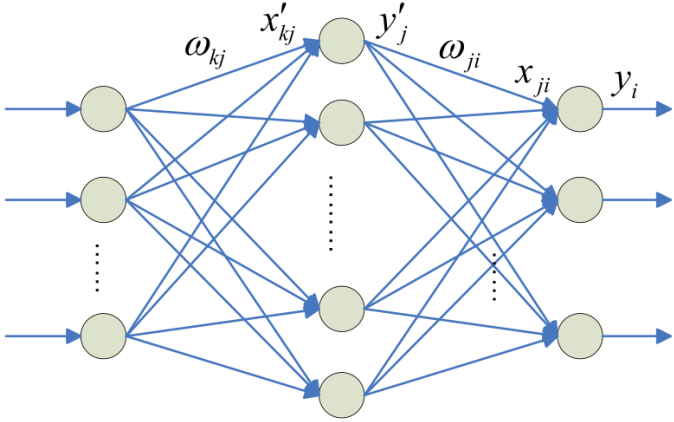

**Figure 2.** Three layer BP Network

A three layer BP network can represent any function in theory. Each neural cell consists of two kinds of functions, the weight function and the activation function. For the hidden layer, we use linear weighting function which simply find the summation of the product of the input and the weights.

$$x'_{kj} = \sum_{i=1}^{n} w_{kj} v_k - b, for all k$$

where b is the bias and $w_{kj}$ is the weight between cell $k$ and cell $j$, $v_k$ is the input $k$. For the activation function, we use sigmoid function which is as follows.

$$y_j = sigmoid(x'_{kj}) = \frac{1}{1 + e^{\sum_{i=1}^{n} w_{kj} v_k - b}}, for all k$$

where sigmoid(x) is $\frac{1}{1+e^x}$.

As for the output layer, the summation function and the activation function is respectively the same as the hidden layer. After constructing the BP network, we use back propagation learning algorithm [19] to modify the weights to get better performance.

**Step 1:** Define the loss function. We use the square error function as our loss function:

$$e(w) = \frac{1}{2} \sum_{i=1}^{n} (d_i - y_i)^2, for all i$$

where L is the final loss, $d_i$ is the desired output and $y_i$ is the real output. Our objective function is:

$$w^* = arg \min_{w} e(w)$$

where $w^*$ is the final weight.

**Step2:** Use gradient descent to modify the weights between the output layer and the hidden layer. The structure of gradient descent for output layer is:

$$w = w - \eta \frac{\partial e}{\partial w}$$

where $\eta$ is the learning rate. To get $\frac{\partial e}{\partial w}$, we use the chain rule with intermedia variables $y_i$ and $\sigma_i$:

$$\frac{\partial e}{\partial w_{ji}} = \frac{\partial e}{\partial y_i} \frac{\partial y_i}{\partial \sigma_i} \frac{\partial \sigma_i}{\partial w_{ji}}$$

where $y_i$ is the generated output and $\sigma_i$ is the summation of the product of w and x:

$$\sigma_i = \sum w_{ji} x_{ji}$$
$$y_i = \frac{1}{1 + e^{\sigma_i}}$$

Then we get:

$$\frac{\partial e}{\partial w_{ji}} = \eta x_{ji} y_i (1 - y_i)(y_i - d_i)$$

As a result, the learning function for the weights between hidden layer and output layer is as follows:

$$w_{ji} = w_{ji} - \eta x_{ji} y_i (1 - y_i)(y_i - d_i)$$

**Step 3:** Back propagate the error to the hidden layer:

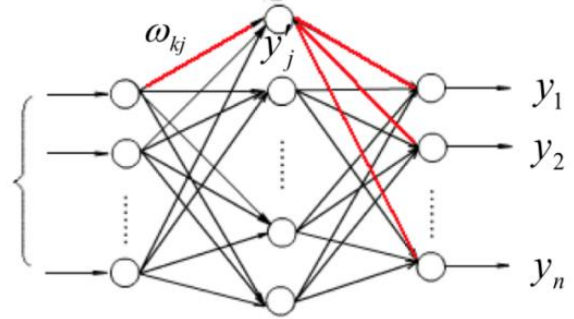

**Figure 3.** Back propagate the error to the hidden layer

By using the chain rule iteratively, we can back propagate the error to the weights between input layer and hidden layer:

$$\frac{\partial e}{\partial w_{ki}} = \sum_{i=1}^{n} \frac{\partial e}{\partial y_i} \frac{\partial y_i}{\partial \sigma} \frac{\partial \sigma}{\partial y'_j} \frac{\partial y'_j}{\partial \sigma'} \frac{\partial \sigma'_j}{\partial w_{kj}}$$

where $\frac{\partial e}{\partial y_i} \frac{\partial y_i}{\partial \sigma} \frac{\partial \sigma}{\partial y'_j}$ is $w_{ji} y_i (1 - y_i)(y_i - d_i)$. Respectively substitute the elements of the equation, $\frac{\partial e}{\partial w_{ki}}$ can be transformed to a form which is more easy to compute and understand:

$$\begin{cases} \dfrac{\partial e}{\partial y_i} = (y_i - d_i) \\ \dfrac{\partial y_i}{\partial \sigma_i} = y_i(1 - y_i) \\ \dfrac{\partial \sigma_i}{\partial y'_j} = w_{ji} \\ \dfrac{\partial y'_j}{\partial \sigma'} = y'_j(1 - y'_j) \\ \dfrac{\partial \sigma'_j}{\partial w_{kj}} = x_{kj} \end{cases}$$

Then the final result is:
$$\frac{\partial e}{\partial w_{ki}} = x_{kj} y'_j (1 - y'_j) \sum_{i=1}^{n} [w_{ji} y_i (1 - y_i)(y_i - d_i)]$$

Also using gradient descent we can derive the learning function:
$$w_{ki} = w_{ki} - \eta x_{kj} y'_j (1 - y'_j) \sum_{i=1}^{n} [w_{ji} y_i (1 - y_i)(y_i - d_i)]$$

**Step 4:** Iteratively using the two equations above to adjust the weights of the neural network until convergence or meet some terminal conditions, we can get the best weights for the model, which means the square error for the desired output and the real output reaches minimum. The process can be demonstrated as **Figure 4**.

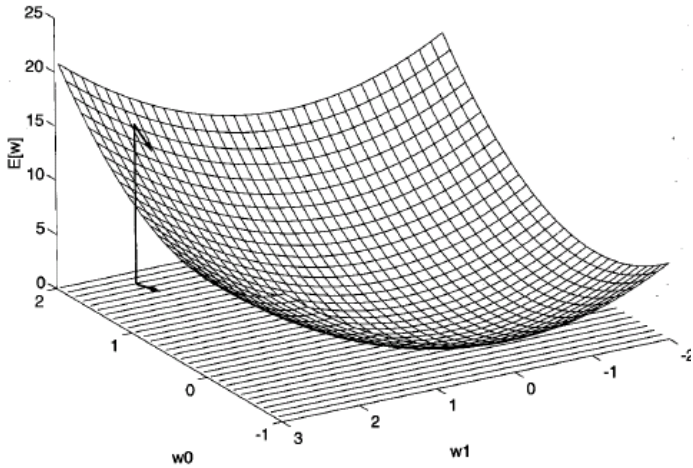

**Figure 4.** The process of gradient descent. The plain is the solution space and the surface above it is the evaluation of the solution. The core idea of gradient descent is to search the solution space according to the gradient[20] to make the loss function reaches the minimum.

During the implementation of the algorithm, the number of cells in the hidden layer is set to 10, cells in the input layer is set to five and cells in the output layer is set to 1. The final output is the evaluation of the region. The label of the training data is collected from Fragile State Index of the year 2016 and the training data itself is collected from the world bank of year 2016. The labels are normalized between 0 and 100 and the label above 80 is considered fragile. The label between 50 and 80 is considered vulnerable and the label below 50 is considered stable. The training result is as follows:

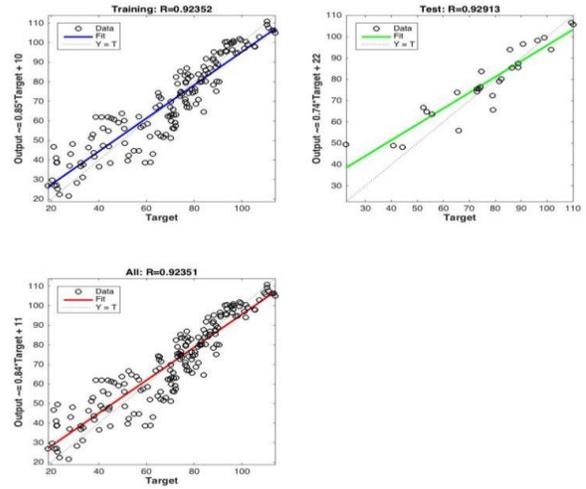

**Figure 5.** Regression Result of the Network

Using BPNN to stand for the network, we can define the Region Stability(RS):
$$RS = \frac{100}{BPNN(x)} - 1$$

where $x$ is a five dimensional vector with each element standing for one particular value of the five indexes. After the operation we can see that RS below 0.25 is considered fragile, between 0.25 and 1 is considered vulnerable and above 1 is considered stable.

## 3. EXPERIMENTS AND DISCUSSION

In this part, we choose three countries that are listed as the most fragile countries in the world on http://fundforpeace.org/fsi/ to implement our model, which are Sudan, Haiti and Somalia.

### 3.1 Analysis of Sudan

In this section, we choose Sudan from the top 10 fragile countries. Sudan is the 5th most fragile country on the list of which the summation of the 12 indexes is more than 100. It is located in Northeast Africa. The basic climate feature of it is hot and dry. In fact, it is one of the hottest countries in the world. Its capital Khartoum is said to be "the world furnace" with an annual average temperature above 30 ℃. And its economy is built on agriculture and husbandry. Army expending is relatively high and the public support is low. The five indexes of Sudan are as follows:

Table 3. Basic Information about Sudan

| LAP | AAT | FO | AMS | PSR | RS |
|---|---|---|---|---|---|
| 250mm | 27 ℃ | 2.765 | 2,89% | 15% | -0.0825 |

We can see that the RS of Sudan is very low. The main factor is the extreme weather of Sudan. The annual average temperature is pretty high while the long term average precipitation is relatively small, which means dry days in Sudan are normal. As a result, the food supply will become a big problem. According to UNDP(United Nations Development Program), the food structure of Sudan is simple and dull. Vegetables are hard to find. The staple food is potato and chicken plus some fruit. What's more, Public Support Rate is relatively low, which means that once the food supply pauses, people will probably rebel against the government.

*3.2 Analysis of Haiti*

Haiti is the 12th most fragile country on the list, which is also one of the poorest countries in the world. Its main economic form is agriculture and its infrastructure is outdated. The climate there varies with latitude with the North being a tropical rain forest climate and the South being a tropical steppe climate. The precipitation is abundant and the average temperature is high. Despite the great climate for plant growth, it suffers a lot from the fragile political state. There are already 104 political parties registered and the government there is weak. The basic information for Haiti is as follows:

**Table 4.** Basic Information about Haiti

| LAP | AAT | FO | AMS | PSR | RS |
|---|---|---|---|---|---|
| 1440mm | 24.7 ℃ | 0.84 | 0.092% | 28% | -0.036 |

Then we implement our model on it. The output result is -0.036, which means it is very dangerous to live in Haiti because it is pretty fragile. The politic reason for the result is that the Annual Military Spending and the Public Support Rate is relatively low. As for the climate factors, long term of high temperature creates a great environment for bacterial to breed. Death rate caused by bacterial infections is pretty high. Despite the inadequate medical situation, the food supply is another emergent factor. 75% of Haiti's population is farmer while the domestic food production cannot meet the need. More than 72% of the food relies on import, which is to say, if the LAP drops, the food supply situation will be more dangerous and the RS will drop sharply.

In summary, long term average precipitation influences the food supply and the annual average temperature has an effect on medical situation. In Haiti, such an undeveloped country, the two factors are even more significant. Respectively set the AAT to 30 and the LAP to 0, the evaluation result is 0.0245 and -0.0386, from which we can see that the region becomes more unstable.

Consistency indexes are calculated by the third layer pairwise comparison matrix. All experimental results pass the test.

*3.3 Analysis of Somalia*

Somalia is the least developed country in the world. Since 2013, the economic status of the country has been very serious. There are 8.2 million hectares of arable land in Somalia, accounting for 13% of the country's land area, while only 100,000 hectares(1.2%) of arable land have been cultivated. In 1990, the agricultural output value accounted for about 20% of its GDP, and the agricultural population accounted for 30% of its total population.

Livestock husbandry is the major economic pillar of Somalia and Somalia is one of the countries with the most per capita livestock in the world. About 80% of its population lives off livestock and semi-agricultural husbandry. The output value of animal husbandry accounts for about 43% of its GDP, and the export income of animal husbandry products accounts for more than 60% of the total export revenue.

As the economy of Somalia has a great deal with Agriculture and Livestock husbandry, it is more sensitive to the climate. Unfortunately, Somali is a country with Tropical desert climate and Savannah climate, which means the annual temperature is high, up to 28 ~ 30 ℃, with the annual rainfall of only 200 to 300mm. This climate results in poor vegetation coverage and extremely serious soil erosion. And finally the reason why Somalia is the least developed country in the world can be explained as the unbalance between reliance on climate and poor climate condition.

The detailed information of the five indexes of Somalia is as follows:

**Table 5.** Basic Information about Somali

| LAP | AAT | FO | AMS | PSR | RS |
|---|---|---|---|---|---|
| 1630mm | 26 ℃ | 0.76 | 0.089% | 35% | -0.056 |

As we can see, the current state of Somali is not optimistic. Due to the extreme weather and the war pressure around the country, it appears to be worse in the following years.

*3.4 Predictions and compares*

In this part, we use Linear Regression[21] to make a prediction of the following states of the countries mentioned above. Linear Regression is widely used to make predictions. It mainly creates a straight line that fits the dataset best. Considering the climate situation in a region usually changes steadily. the linear regression will fit the data well. The dataset we use consists of ten years of the environmental and political states of the top 10 fragile countries on http://fundforpeace.org/fsi/ of 2017 from World bank Database between 2010 to 2017. They are shown in **Table 6~ Table 8**.

We concentrate on the three countries we mentioned. Before making predictions, we use our model to evaluate the previous RS. The result is shown in **Table 6~ Table 8**.

**Table 6.** States in Sudan

| year | 2010 | 2011 | 2012 | 2013 | 2014 | 2015 | 2016 | 2017 |
|---|---|---|---|---|---|---|---|---|
| LAP/mm | 239 | 270 | 265 | 268 | 258 | 253 | 250 | 255 |
| AAT/ | 25 | 26 | 24 | 25 | 24 | 24 | 28 | 27 |
| FO | 2.765 | 2.770 | 2.768 | 2.76 | 3.6 | 2.765 | 2.765 | 2.765 |
| AMS | 2.85% | 2.83% | 2.78% | 2.75% | 2.89% | 2.82% | 2.86% | 2.81% |
| PSR | 17% | 16% | 13% | 16% | 17% | 18% | 12% | 13% |
| RS | -0.0825 | -0.08 | -0.0830 | -0.0835 | -0.0973 | -0.0875 | -0.102 | -0.127 |

**Table 7.** States in Haiti

| year | 2010 | 2011 | 2012 | 2013 | 2014 | 2015 | 2016 | 2017 |
|---|---|---|---|---|---|---|---|---|
| LAP/mm | 1370 | 1440 | 1440 | 1440 | 1440 | 1440 | 1440 | 1440 |
| AAT/℃ | 24.3 | 24.6 | 23.5 | 23.7 | 23.8 | 24.4 | 24.8 | 24.6 |
| FO | 0.83 | 0.85 | 0.86 | 0.85 | 0.84 | 0.82 | 0.85 | 0.86 |
| AMS | 0.090% | 0.093% | 0.091% | 0.082% | 0.087% | 0.088% | 0.080% | 0.076% |
| PSR | 29% | 30% | 27% | 29% | 28% | 28% | 27% | 27% |
| RS | -0.0354 | -0.037 | -0.0362 | -0.0378 | -0.036 | -0.0388 | -0.040 | -0.042 |

**Table 8.** States in Somalia

| year<br>index | 2010 | 2011 | 2012 | 2013 | 2014 | 2015 | 2016 | 2017 |
|---|---|---|---|---|---|---|---|---|
| LAP/mm | 1530 | 1530 | 1570 | 1530 | 1630 | 1630 | 1630 | 1630 |
| AAT/ | 25 | 27 | 26 | 26 | 27 | 26 | 25 | 27 |
| FO | 0.76 | 0.35 | 1.56 | 1.33 | 1.32 | 0.85 | 0.88 | 1.34 |
| AMS | 0.093% | 0.088% | 0.103% | 0.087% | 0.076% | 0.089% | 0.086% | 0.085% |
| PSR | 36% | 37% | 35% | 33% | 35% | 35% | 39% | 32% |
| RS | -0.056 | -0.057 | -0.043 | -0.044 | -0.042 | -0.039 | -0.032 | -0.030 |

Then we use Linear Regression to make a prediction of RS of the following five years. Before that we make use of the correlation coefficient to indicate the linear relativity between year and RS. The results are -0.8265,    -0.8689, 0.9547. We can see that they are linear relative. And then we operate linear regression and the result is shown in Figure 5.

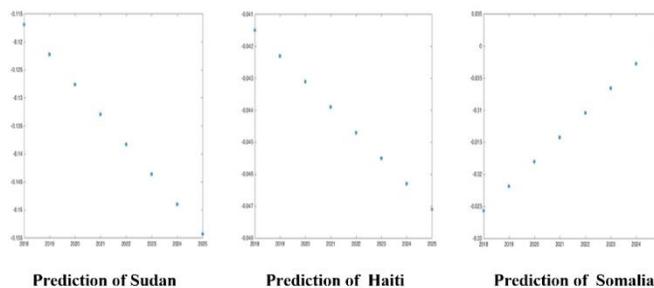

**Figure 5.** Predictions of RS

### 4.Conclusions and policy implications

The prediction results show that Somali's RS index has a positive prospective in the future. Compared with Sudan and Haiti, Somalia is a relatively homogeneous religious country. At present, it has a population of more than 7 million, with somalis, the main ethnic group, make up about 97 percent. Due to historical and geographical factors, Somali people are still kept in the late primitive society of tribalism and remain a tribal tradition that patriarchal traces the origin of the custom. Different from Somalia, Sudan and Haiti both suffered a lot from warlord chaos, which explains why they will have a poor development in the future.

Actually, the developments of many poor countries have been trapped in a vicious circle, that poverty leads to more environmental damaging behaviors(like cutting trees, polluting water, etc), which then leads to severe soil erosion resulting in decreasing crop yields and finally intensify poverty.

The analytical model has been implemented both on developing countries and developed countries. Results show that climate change plays a more influencial role in regional stability of developing countries (like Somali, Haiti, etc) than developed countries. According to the estimation of the World Bank, agriculture constitutes about 45% to 55% of the GNP (gross national product) in smallholder plantation dominating country. While in developed countries, with high agricultural productivity, the figure is less than 3%. This phenomenon indicates that some poorest developing countries can learn from developed countries and take measures to get rid of poverty.

Based on the analysis above, the market-based developed countries tend to be more stable under bad climate conditions. We propose a vision for the development of the poor developing countries that people in those countries need to adopt a market-based agricultural system rather than "self-sufficient farming" to pursue a better life in the future. By breaking restrictions on imports, expanding farms, putting a large number of personnel engaged in administration in agricultural production, and increasing the level of mechanization, farms will produce more product to the market. Then the competition between farms will increase productivity. We need to recognize that only through the break of small-scale peasant economy, can it be possible to form a market-oriented agriculture system, which will improve people's life quality and finally increase the stability of a region.